\pdfoutput=1
\documentclass[%
 reprint,
 amsmath,amssymb,
 aps,
]{revtex4-1}

\usepackage{natbib}
\usepackage{color}
\usepackage{graphicx}
\usepackage{dcolumn}
\usepackage{bm}
\usepackage{hyperref}
\usepackage{amsmath}

\begin{document}

\preprint{APS/123-QED}

\title{Elastocapillary driven assembly of particles at free-standing smectic-A films}

\author{Mohamed Amine Gharbi$^{1}$}
\email{Corresponding author: mohamed.gharbi@umb.edu}
\author{Daniel A. Beller$^{2}$}
\author{Nima Sharifi-Mood$^{3}$}
\author{Rohini Gupta$^{3}$}
\author{Randall D. Kamien$^{4}$} 
\author{Shu Yang$^{5}$}
\author{Kathleen J. Stebe$^{3}$}%
 \email{Corresponding author: kstebe@seas.upenn.edu}
\affiliation{%
$^1$Department of Physics, University of Massachusetts Boston, Boston, MA 02125, USA
}%
\affiliation{%
$^2$School of Engineering, Brown University, Providence, RI 02912, USA
}%
\affiliation{%
$^3$Department of Chemical and Biomolecular Engineering, University of Pennsylvania, Philadelphia, Pennsylvania 19104, USA
}%
\affiliation{%
$^4$Department of Physics and Astronomy, University of Pennsylvania, Philadelphia, Pennsylvania 19104, USA
}%
\affiliation{%
$^5$Department of Materials Science and Engineering, University of Pennsylvania, Philadelphia, Pennsylvania 19104, USA
}%

\date{January 5, 2018}

\begin{abstract}
Colloidal particles at complex fluid interfaces and within films assemble to form ordered structures with high degrees of symmetry via interactions that include capillarity, elasticity, and other fields like electrostatic charge.  Here we study microparticle interactions within free-standing smectic-A films, in which the elasticity arising from the director field distortion and capillary interactions arising from interface deformation compete to direct the assembly of motile particles.  New colloidal assemblies and patterns, ranging from 1D chains to 2D aggregates, sensitive to the initial wetting conditions of particles at the smectic film, are reported. This work paves the way to exploiting LC interfaces as a means to direct spontaneously formed, reconfigurable, and optically active materials. 

\end{abstract}

\pacs{Valid PACS appear here}
\maketitle


\section{Introduction}

Colloidal particles can self-assemble via weak interactions of order $k_{\mathrm{B}}T$ to form structures influenced by particle shape \cite{ref1}, patchiness \cite{ref2}, and entropic interactions \cite{ref3}. Alternatively, colloids can assemble under the direction of external inputs such as electromagnetic fields to form structures strongly influenced by the fields themselves \cite{ref4,ref5}. Other strategies to direct colloid assembly exploit fields that occur within soft matter, including capillary interactions \cite{ref6,ref7} and elastic forces \cite{ref8,ref9}. These fields can impose rich anisotropic potentials that direct the formation of new structures with potential for use in generation of new functional materials \cite{ref10,ref11}. 

\begin{figure}
\includegraphics{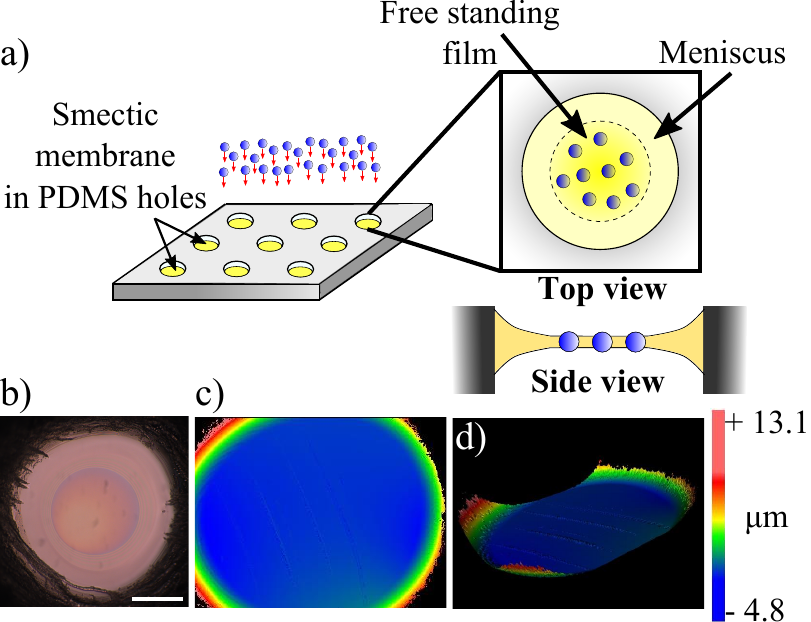}
\caption{Free-standing smectic-A (SmA) films. (a) Trapping of solid particles at smectic membranes prepared in millimetric pores in a PDMS sheet. (b) Optical microscopy image of a smectic film in a pore. The inner circle shows the free-standing film, and the outer one indicates the pore meniscus around it (scale bar: 500 $\mu$m). (c) Topographic profile of the air/smectic interface. (d) The 3D depiction of the profile shows that the free-standing film is flat, and its thickness is around $l \approx 3.5 \pm 0.5$ $\mu$m.}
\label{Fig1}
\end{figure}

There is a strong interest in exploiting elastic energy fields in liquid crystals (LCs) as a means of directing assembly \cite{ref12}. The ability of LCs to organize particles into regular structures provides novel routes to control ordering transitions, with the possibility of building complex artificial structures for diverse applications. Since LCs are readily reconfigurable, they offer opportunities to make responsive devices including smart energy efficient windows \cite{ref13}, responsive optical components \cite{ref14}, and sensors \cite{ref15}. Smectic LC films, in particular, offer important degrees of freedom that can be used to direct particles into new structures \cite{ref16,ref17,ref18,ref19,ref20}, owing to elastic energy and the tendency of smectic layers to organize into focal conic domains \cite{ref21,ref22,ref23,ref24}.

There is also a growing interest in investigating the behavior of particles captured at fluid interfaces from both fundamental and technological points of view \cite{ref25}. Conceptually, colloidal interactions at fluid interfaces have important analogies throughout classical and statistical physics ranging from Casimir interactions \cite{ref26} to interactions of protein inclusions in lipid bilayers \cite{ref27}. Improved understanding of these interactions, therefore, provides insight into these related systems \cite{ref27B}. Practically, these interactions are harnessed to develop strategies for novel material fabrication, which find relevance in many applications including Pickering emulsions \cite{ref28}, foams and coatings \cite{ref29}, and permeability control of fluxes across interfaces \cite{ref30}. While colloidal interactions at interfaces of isotropic liquids are now well understood, this is not the case for colloids at interfaces of complex ordered fluids, especially LC phases \cite{ref31,ref32,ref33,ref34,ref35,ref36}.

LC interfaces are especially promising sites for assembly, since they offer both capillarity and elasticity to guide the formation of ordered structures. Studies of the dynamics of rod-like particles at the free surface of thin hybrid nematic films \cite{ref37} reveal complementary roles for elasticity and capillarity on the assembly of cylindrical particles. For particles trapped in thin nematic films, long-range interactions mediated by elastocapillarity occur via the formation and interactions of novel defect configurations \cite{ref38,ref39}. While these recent works focus on capillarity at nematic interfaces, little is known about capillary effects at more complex LC phases such as smectic films. An  overview of experimental and theoretical progress on understanding colloidal interactions in liquid crystals, such as nematic and smectic phases, can be found in \cite{ref39B, ref39C}.

In this work, we explore elastocapillary interactions at smectic interfaces by studying the assembly of micrometric beads at free-standing smectic-A (SmA) films. Most of the previous studies, presented in the review by Bohley and Stannarius \cite{ref16}, focus on the assembly of fluid droplets in free-standing smectic films. Here we choose to work with solid particles in order to control their anchoring properties via surface treatment but also to avoid any shape deformation that could be induced by elastic or capillary effects. In these films, microbeads interact through capillarity controlled by their wetting properties and via elasticity induced by the distortion of smectic layers around them. The resulting potential drives the assembly of particles into several novel structures depending upon their wetting configurations. These structures are mobile and migrate towards the menisci that form at the edges of the pores that contain the free-standing SmA films.

\section{Experimental Section}

\begin{figure}
\includegraphics[scale=1.4]{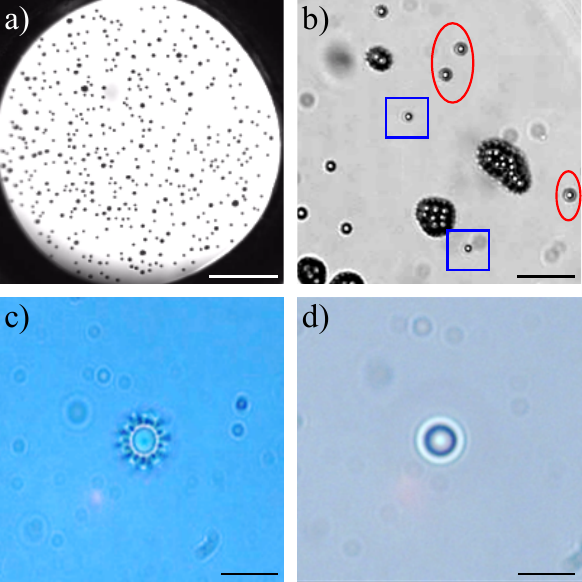}
\caption{Solid microparticles at free-standing smectic films. (a) Bright-field image of a high density of isolated particles captured on a free-standing SmA film (scale bar: 500 $\mu$m). The membrane with particles is stable for many hours. (b) Higher magnification of particles at the smectic film (scale bar: 50 $\mu$m) showing the coexistence of particles with a corona of FCDs (in red ellipses) and without the corona (in blue squares). (c) Optical  microscopy image of an individual particle with a corona of defects similar to the flower texture. (d) Optical microscopy image of a particle without the corona. (Scale bars in (c) and (d): 10 $\mu$m).  }
\label{Fig2}
\end{figure}

Silica beads of nominal diameter $2R$ = 5 $\mu$m (Polysciences, Inc.) are trapped at the interface between air and free-standing SmA films. The SmA membranes are formed by spreading a small amount of 4-n-octyl-4'-cyanobiphenyl (8CB purchased from Kingston Chemicals Limited that displays a SmA phase between $23.3\,^{\circ}\mathrm{C}$ and $33.4\,^{\circ}\mathrm{C}$) across 2 mm diameter pores in a film of cross-linked polydimethylsiloxane (PDMS from Sigma-Aldrich) using a sharp-edged blade. The elastomer and curing agent are mixed at a ratio of 10 parts to 1 (10:1). The mixture is then set under vacuum until almost all the bubbles disappear from the solution. Thereafter, we place the solution into the oven for about an hour at $80\,^{\circ}\mathrm{C}$ until the PDMS becomes solid.

The thickness of the smectic film is obtained by measuring the vertical positions of both upper and lower interfaces of the free-standing film using the technique of scanning white-light interferometry (SWLI, Zygo NewView 6200 interferometer). The shape of the menisci around beads is determined using the technique of single wavelength interferometry (Newton Rings). The surface of the silica beads is modified by the covalent attachment of a monolayer of N,N-dimethyl-N-octadecyl-3-aminopropyl trimethoxysilyl chloride (DMOAP from Sigma-Aldrich) following Ref. \cite{ref32} to induce strong homeotropic anchoring. The SmA film also has strong homeotropic anchoring in contact with air at both interfaces. The corresponding experimental system is sketched in Fig.\ref{Fig1}-a.

The system was studied under an upright optical microscope (Zeiss AxioImager M1m) in transmission mode and a set of crossed polarizers. Images were recorded with a high-resolution camera (Zeiss AxioCam HRc) and high-speed camera (Zeiss AxioCam HSm). Particle trajectories were determined using standard tracking procedures (ImageJ, National Institute of Mental Health, Bethesda, MD, USA).

\section{Results and Discussion}

\begin{figure}
\includegraphics[scale=0.90]{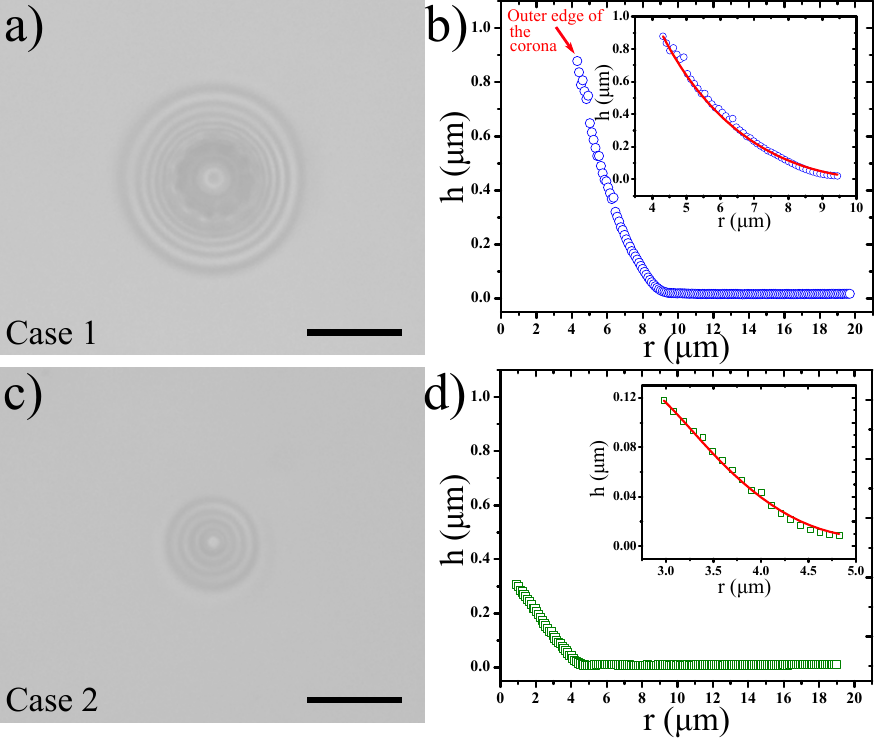}
\caption{Profile of the smectic meniscus around individual particles. (a) Interferometric image of a particle with a corona of FCDs in the reflection mode (case 1). The fringes indicate the formation of a meniscus around the colloid but could not be observed inside the corona. (b) The corresponding meniscus profile. The arrow shows the outer radius of the corona of FCDs. (c) Interferometric image of a particle without the corona (case 2) and (d) its corresponding meniscus profile. (Scale bars 10 $\mu m$). The solid lines in the insets of (b) and (d) represent the corresponding fit of the menisci with the expression given by the Eq. 2. }
\label{Fig3}
\end{figure}

Free-standing SmA films were prepared at room temperature with a thickness in the range of $l$ = 3.5 $\pm$ 0.5 $\mu$m, thinner than the dimension of solid particles (nominal diameter $2R$ = 5 $\mu$m). The thickness of the film can be controlled by changing both the amount and drawing speed of SmA over the pores. The resulting films are flat with edges surrounded by menisci as shown in Fig.\ref{Fig1}-b, Fig.\ref{Fig1}-c, and Fig.\ref{Fig1}-d. These menisci stabilize films by serving as material reservoirs and provide connections between flat regions and edges of pores in the supporting PDMS sheet \cite{ref40}. The free-standing films are composed of 2D equidistant stacked layers, where the thickness of each layer is around $a$ $\approx$ 3.5 nm \cite{ref35}. The smectic is in contact with air at both upper and lower interfaces where it interacts via the interfacial tension $\gamma_{SA} \sim 10^{-2}$ N/m that reduces the fluctuations of the SmA at interfaces as described in Ref. \cite{ref41,ref42}. 

\begin{figure*}
\includegraphics{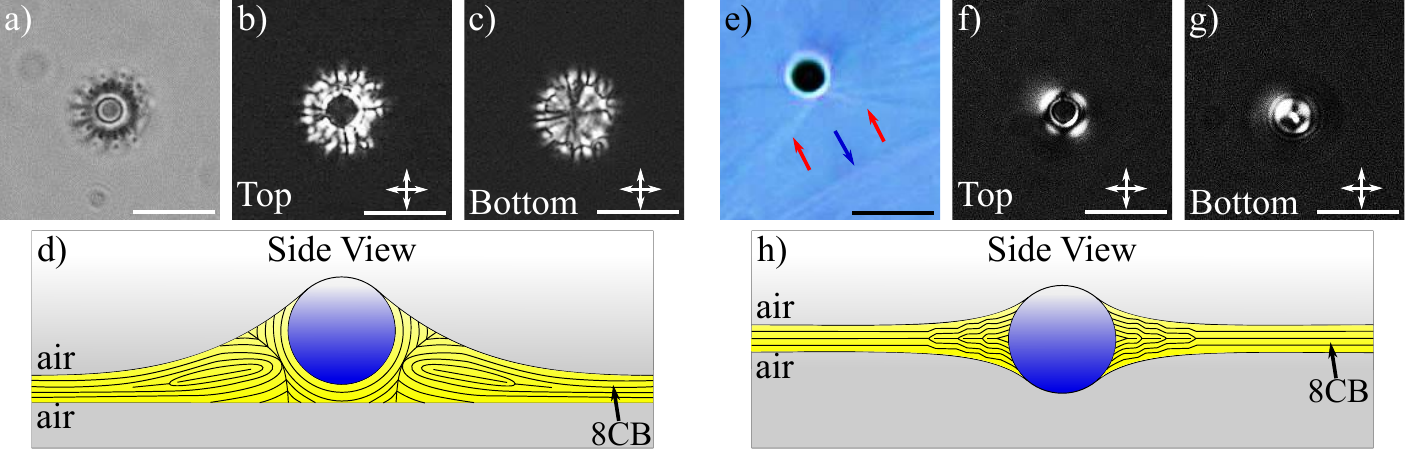}
\caption{Possible wetting configurations of solid particles at a free-standing SmA film. Bright-field images of a particle surrounded by a corona of focal conic domains like the flower texture (a), and a particle without the corona (e). (b, c, f, and g) Polarized optical microscopy (POM) images of particles in both configurations. (b, c) The images confirm that the particle is trapped by the upper interface. (f, g) The images show that the particle is trapped by both interfaces (upper and lower). (Scale bars: 15 $\mu m$).  (d, h) Schematic illustrations of the corresponding cross section textures of the smectic layers around a particle decorated by the corona (d) and a particle without the corona (h). (Scale bars: 15 $\mu$m). }
\label{Fig4}
\end{figure*}

Once the films are formed, we cover the PDMS pores with a coverslip and wait for a few hours until the films are stabilized. Particles are then gently aerosolized in a closed container with compressed air. Once particles are airborne, the substrate supporting the free-standing films is introduced, and particles are effectively deposited via gravity with minimal aggregation as shown in Fig.\ref{Fig2}-a \cite{ref32}. The beads are then trapped at the air/SmA interface due to their strong binding energy as discussed in \cite{ref43} (trapping energy $\Delta E$ $>$  $10^8\,k_{\mathrm{B}}T$) \cite{ref52}. This process is very delicate because the elastic membrane may rupture in some cases. However, once the beads are trapped, the films are very stable. In some cases, we observe that particles seem to change their size slightly over time. This is because they are sinking in the smectic film till their wetting properties are satisfied. We measure a contact angle of our particles at the air/SmA interface to be around $\theta \sim 30 ^{\circ}$, obtained macroscopically by measuring the contact angle of 8CB droplets on silanized glass slides. For individual particles, we distinguish two configurations depending on the texture of the smectic around them. In most instances, particles are decorated with a corona of focal conic domains (FCDs) (Fig.\ref{Fig2}-c); we refer to this scenario as case 1. However, we also observe particles that lack this corona (Fig.\ref{Fig2}-d), which we refer to as case 2. Both configurations can coexist as shown in Fig.\ref{Fig2}-b. This is not the case when non-treated particles are used instead. The particles don't form coron$\ae$. In this work, we focus on the behavior of particles with strong homeotropic anchoring. As described in detail below, they interact with each other in a manner dependent on their trapped state, and on the smectic structures which form around the individual beads.

The interferometric results presented in Fig.\ref{Fig3} show that solid particles captured at smectic films deform the interface around them. The curvature of the interface is controlled by particle wetting properties at the contact line between air, smectic, and particle, influenced by the smectic layers. The resulting shapes of the smectic menisci differ significantly from the distortion around a particle in a thin film of isotropic fluid, which corresponds simply to a monopole in polar coordinates centered at the particle center of mass. Rather, the observed menisci decay rapidly with distance from the particles and end abruptly some distance from the colloid. This is shown in Fig.\ref{Fig3}-b and Fig.\ref{Fig3}-d, respectively, for particles with the corona (indicated by subscript 1) and particles without the corona (indicated by subscript 2). There are distinct differences in menisci near the particles for the two cases, including the height of the smectic meniscus profile above the flat film near contact with the particle ($h_{0,1} \approx 0.87$ $\mu$m versus $h_{0,2} \approx 0.3$ $\mu$m) and the distances at which the distortions vanish ($r_{1,1} \approx 9.5$ $\mu$m from the center of the colloid versus $r_{1,2} \approx 4.72$ $\mu$m), indicating that  colloids surrounded by FCDs are located at higher vertical positions. These observations suggest that these particles assume two different wetting configurations; particles can either be trapped by the upper interface of the film, as we infer for particles with corona (Fig.\ref{Fig4}-d), or they can span both interfaces, as we infer for particles lacking corona (Fig.\ref{Fig4}-h). We confirm these states by analyzing the birefringence patterns of the SmA around isolated particles under crossed polarizers using optical microscopy. Fig.\ref{Fig4}-c reveals the presence of a few layers of SmA under the colloid decorated by the corona of FCDs. However, this is not the case for the colloid without the corona as shown in Fig.\ref{Fig4}-g. We also find that the corresponding fluorescence images shown in the supplementary materials (SM Fig.1), are in agreement with the results obtained by polarized optical microscopy (see section ``Fluorescence microscopy'' for details). The mechanism that controls the wetted area of particles depends on whether the colloid can penetrate all smectic layers or some of them as explained in Ref. \cite{ref35}. Unfortunately, it is difficult to control experimentally the adsorption state of particles at free-standing smectic films. 

Our rather gentle method of using an air spray to capture particles favors the case with the corona of FCDs ($\sim 90\%$), perhaps because of energy barriers associated with the penetration of the smectic layers. As we mentioned previously, however, we always observe both cases in the same sample. Further attempts to control the process of trapping particles without the corona using a strong air spray were not successful, as the stronger air burst ruptures the smectic membranes.

The configuration with the corona of FCDs in a free-standing smectic film differs from other coron$\ae$ configurations in the literature. Previously observed coron$\ae$ in smectic films are comprised of smectic wrinkles, as reported in Refs. \cite{ref35,ref42,ref44}, unlike the ring of FCDs in our case. The FCDs shown here decorate the particle like the petals of a flower (Fig.\ref{Fig4}-a), and the resulting structures are very similar to the smectic flowers that form in droplets \cite{ref45} and around particles at solid substrates as described in our previous works \cite{ref22,ref23,ref24}. In particular, based on the geometry of the smectic film near the colloid, the FCDs' hyperbolic focal curves are expected to bend radially outward from the colloid. Fig.\ref{Fig4}-d presents a schematic illustration of a cross-section of this configuration, in which the hyperbolic focal curves of two FCDs are  oriented outward from the particle and span between the two smectic-air interfaces. The illustration shows the smectic layers near the particle as spheres concentric about the particle center, in accordance with strong homeotropic anchoring at the particle's surface. It is not clear precisely how the curved layers of the FCDs join smoothly onto the texture around the particle in three dimensions, although the system resembles the previously studied smectic flower texture \cite{ref45,ref22,ref23,ref24}.

The presence of particles in free-standing films promotes local deformation of the smectic interface. This deformation is controlled by wetting conditions imposed via functionalization of particles with DMOAP, although contact line pinning, widely reported for colloids at isotropic fluid interfaces, probably plays a role \cite{ref53}. Thus, to satisfy the condition of a contact angle around $\theta \sim 30 ^{\circ}$ at the air/SmA interface, the particles create a meniscus around them. However, for smectic LCs the formation of menisci is different from the case of isotropic fluids because of their lamellar structure. In fact, the smectic LC forms elementary edge dislocations to increase the number of layers by one or more and vary the thickness around particles, as explained in previous works \cite{ref41,ref46,ref47}. This effect implies that the density of dislocations is then correlated to the curvature of the meniscus.

Based on previous experimental works and theoretical analyses performed on SmA films, we distinguish three regions of the SmA meniscus: near the particle where the density of dislocations is high, away from the particle where the density of dislocations is medium, and far from the particle where the density of dislocations is low \cite{ref44,ref47}. In the region where the deformation of the interface is very pronounced, it has been shown that the dislocations are discrete and packed together. These dislocations interact with each other and form giant dislocations of very large Burgers vectors (up to one hundred layers). However, the resulting dislocations are unstable with respect to the formation of focal conics and turn into a ring of \textquotedblleft oily streaks\textquotedblright  \cite{ref48}. This explains the origin of the corona around particles in our first configuration (Fig.\ref{Fig4}-a). In the region where the deformation of the interface is less pronounced, the density of dislocations is medium, and the typical distance between them is smaller than the size of distortions at the free surface induced by a single dislocation, but large enough to prevent the formation of giant dislocations and focal conics. Since the extent of the region with low dislocation density (known as the vicinal region) is negligible \cite{ref47}, the menisci around particles in both configurations (not including the FCDs in the corona case) should be in the area of medium density of dislocations (Fig.\ref{Fig3}-a,b and Fig.\ref{Fig3}-c,d). In this region, the dislocations are sufficiently closely spaced that they can be approximated by a continuous distribution of infinitesimal dislocations, so that the free surface of the film is smooth, with no apparent steps observed in the height profile. By analogy with the case of a SmA meniscus created by a cylindrical wall, the free energy of the system is given by \cite{ref47}:
\begin{multline}
\mathcal{F}[h(r)]=2 \gamma_{SA} \int_{r_0}^{r_1}2\pi r\mathrm{d}r \left( \sqrt{1+(dh/dr)^2}-1\right)  \\ + 
2 (\gamma_{PS}-\gamma_{PA}) 2\pi r_{0} h(r_0)+2 \Delta p \int_{r_0}^{r_1}2\pi r\mathrm{d}r h(r)\\ - \int_{r_0}^{r_1} 2\pi r\mathrm{d}r E \frac{2}{a}   \frac{dh}{dr}, 
\end{multline}
where $h(r)$ is the height of the meniscus as function of the distance $r$ from the center of particle. $\gamma_{SA}$, $\gamma_{PS}$, and $\gamma_{PA}$ are the smectic/air, particle/smectic and particle/air surface tensions, respectively. $r_0$ is the radius of the contact line where air, particle and smectic meet. For a contact angle $\theta \sim 30 ^{\circ}$ we estimate $r_0$ to be around $r_0 \approx 1.35 \mu$m. In this expression, the first term corresponds to the excess of the surface free energy. The second term corresponds to the change of the surface energy at the particle surface. The third term corresponds to the work of the pressure difference $\Delta p= p_{\mathrm{air}} - p_{\mathrm{smectic}}$ between air and smectic. The fourth term is the energy of dislocations of density $\rho= -(2/a)(dh/dr)$, where $E$ is the core energy of an elementary dislocation and $a$ is the smectic layer spacing. The minimization of $\mathcal{F}$ with respect to $h(r)$ gives the expression for the profile $h(r)$ (see Supplementary Materials for details): 
\begin{multline}
h(r)= h_{\infty}+(ec - r_{\infty})\ln\left[ r/r_{\infty} + \sqrt{(r/r_{\infty})^2-1}\right] \\ + e\left[ \sqrt{r^2-r_{\infty}^2} - \frac{r_{\infty}^2 + c r}{\sqrt{r^2-r_{\infty}^2}} \right] 
\end{multline}
where $h_{\infty}$, $r_{\infty}$ and $c$ are integration constants, and $e$ is a small material parameter given by $e=E/(\gamma_{SA} a)$  \cite{ref47}. Note that the limits of integration $r_0$ and $r_1$ do not appear directly in Equation 2, nor does the term from $\mathcal{F}$ proportional to $(\gamma_{PS}-\gamma_{PA})$. In the use of Equation 2 below, we are therefore free to replace $r_0$, the radius of the contact line, with a new inner radius appropriate for the case at hand. 

The insets of Fig.\ref{Fig3}-b and Fig.\ref{Fig3}-d show the fits of the measured surface profile with the analytical expression of $h(r)$ in both configurations.  The meniscus's measured width, maximum height, and maximum slope are  used to obtain $h_{\infty}$, $c$, and $e$, leaving $r_\infty$ as a single fitting parameter for the shape of the height profile. For Case 1 with the corona, the model is fit to our experimental data with $h_{\infty} \approx 3.18$ $\mu$m, $r_{\infty} \approx 2.11$ $\mu$m, $e \approx 0.24$ and $c \approx -2.42$ $\mu$m in the region outside the corona. The fit in this case is excellent. For Case 2 without the corona, the model fits the meniscus height data poorly if applied to the entire range of distances $r$ between the top of the meniscus and the flat film. It can be seen by eye in Fig.\ref{Fig3}-d that the meniscus height varies roughly linearly with $r$ at small $r$. However, if we restrict our attention to the outer region of the meniscus with $r\gtrsim 3$ $\mu$m, as shown in the inset of Fig.\ref{Fig3}-d, then the model fits well with  $h_{\infty} \approx 1.89$ $\mu$m, $r_{\infty} \approx 1.90$ $\mu$m, $e \approx 0.28$ and $c \approx 0.61$ $\mu$m. Because $e$ is a material parameter, it is important to note that the values of $e$ obtained from Case 1 and Case 2 are in reasonable agreement.  The reason for the roughly linear height profile for $r\lesssim 3$ $\mu$m in Case 2 is unclear, but strong homeotropic anchoring at the colloid likely causes significant distortions near the colloid that are not captured by the model of Ref.~\cite{ref47}, in  which no anchoring is assumed.

In the model described by Eq.~1, the spacing between dislocations is maintained by the interfacial tension $\gamma_{SA}$. An additional expected source of repulsion between the dislocations is the elasticity of the smectic bulk. The curved height profile associated with the arrangement of dislocations implies considerable distortion of the smectic layers at all heights,  as illustrated in Fig.\ref{Fig4}. The elastic free energy cost associated with a small vertical layer displacement field $u(r)$ is given by \cite{ref49}:
\begin{multline}
\mathcal{F}_{\mathrm{elastic}}=\frac{1}{2}\int\mathrm{d}^3 \mathbf{x} \left\lbrace  B \left[ \frac{\partial u(r)}{\partial z}\right] ^2 + K \left[ \bigtriangleup u(r)\right] ^2   \right\rbrace 
\end{multline}
where $B \thicksim$ $10^6$ N/$\mathrm{m}^2$ and $K \thicksim$ $10^{-11}$ N are the elastic constants associated with the compression and bending of the smectic layers, respectively, and $\bigtriangleup$ is the Laplace operator. Unfortunately, adding $\mathcal{F}_{\mathrm{elastic}}$ to $\mathcal{F}[h(r)]$ in Eq.~1 significantly complicates the mathematics, and finding an energy-minimizing displacement field $u(r)$ is outside the scope of this paper, especially when the possibility of FCDs must be taken into account. Qualitatively, however, the energetic cost of smectic elastic distortions in the meniscus plays an important role in the interparticle dynamics that we observe. We note that further contributions to the free energy induced by the adsorption of particles at the free-standing films and the reorganization of smectic molecules near particles can be considered as discussed in Ref. \cite{ref50}.

\begin{figure}
\includegraphics[scale=1.2]{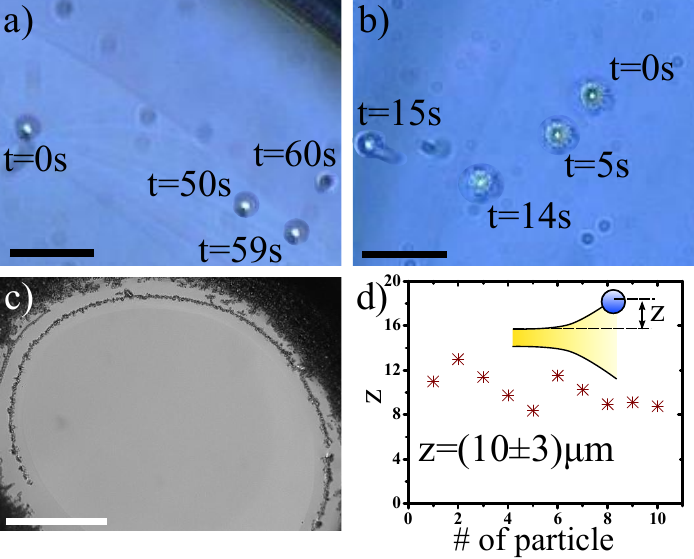}
\caption{Mobile particles at free-standing SmA films. (a) Particles without the corona migrate to the meniscus that surrounds the film. (b) Migration of particles decorated by the FCD corona to the meniscus. (Scale bars: 20 $\mu$m). (c) Ring structure formed by many particles after their migration to the meniscus (scale bar: 500 $\mu$m). (d) Particles fix their vertical position z $\sim$ 10$\pm$3 $\mu$m at the meniscus.   }
\label{Fig5}
\end{figure} 

Once the particles are captured at the membrane, we observe a drainage at the edge of the pore meniscus and the zipping formation of oily streaks. As this drainage occurs, we see a circulating flow in the film that moves colloids close to the meniscus at the edge of the pore. Near the edge, we see strong local capillary interaction when the pore and particle menisci overlap. The particles are then expelled from the film into the meniscus at the pore as shown in Fig.\ref{Fig5}. By  migrating to the  meniscus, the particles reduce the bulk and surface deformations and minimize the excess of the free energy created by elastocapillary effects. The force responsible for expelling particles from the film arises from the interplay between elasticity and capillarity. Once the colloids migrate, we observe under the microscope a relaxation of the smectic layers to their equilibrium configuration. At the pore meniscus, particles fix their zenithal position at a height around $z \thicksim 10\pm 3 \mu$m (Fig. \ref{Fig5}-d) and form a ring around the free-standing film after a few hours ($\thicksim$ 4 h) as shown in Fig. \ref{Fig5}-c.  This position $z$ of particles is determined by their wetting conditions at their surface. The height is fixed where the contact angle of particles at the air/SmA interface is satisfied without creating any interfacial deformation and with minimal bulk distortions. This effect is very similar to the behavior of solid particles trapped in a nematic vesicle with non-uniform thickness reported in \cite{ref50}. In that system, the colloids migrate to a thicker region of the vesicle to satisfy their wetting conditions and minimize capillary distortions at the nematic interface.

\begin{figure*}
\includegraphics[scale=1.05]{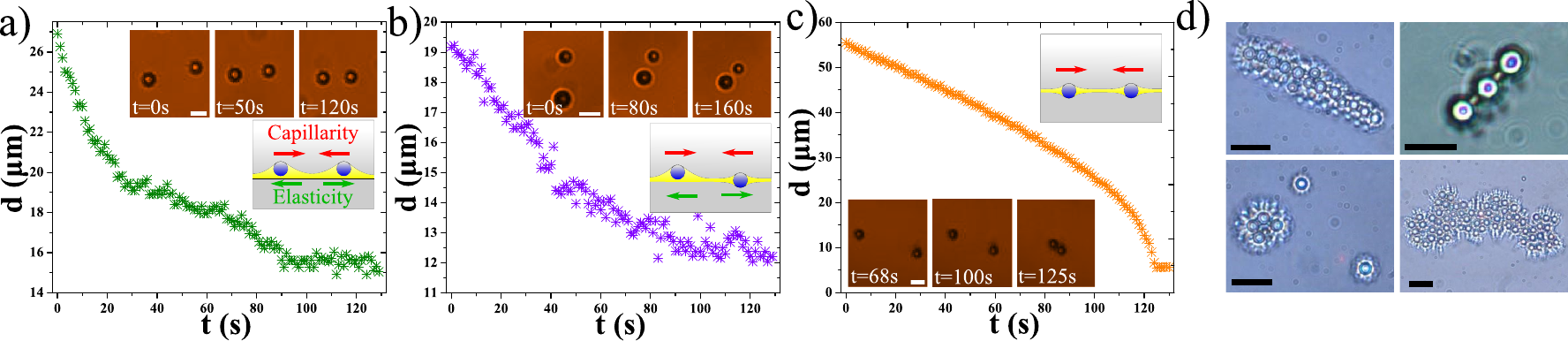}
\caption{Pair interaction and assembly of particles at free-standing smectic films. Time dependence of the separation distance between two particles with FCD coron$\ae$ (a), between a particle with a corona and a particle without a corona (b), and between two particles without coron$\ae$ (c). Insets represent the sequential optical images capturing interactions of particles in each configuration. In the case of particles with coron$\ae$, and the case of one particle with a corona and another without, we observe metastable nonzero equilibrium distances that persist for a few minutes, after which the particles come into contact due to the strong capillary attraction that induces reorganization of the FCDs. (Scale bars: 10 $\mu$m.) (d) Optical microscopy images of the patterns formed by particles at the smectic films. The resulting structures range from 1D chains to 2D aggregates that depend on the wetting configuration of particles. (Scale bars: 15 $\mu$m.}
\label{Fig6}
\end{figure*} 

At the free-standing film, the circulating flow convects colloids toward neighboring particles before they are ejected to the pore meniscus. The particles also create large distortions of the smectic layers that lead to interparticle elastic interactions. These interactions contribute to the attraction between the beads that we observe at large distances. But it is hard to decouple this behavior from the collective motion due to the circulating flows. For this reason, we focus on the local pair interaction at shorter ranges. The colloids interact when they come close enough to each other (center-to-center typical separation distance around $d \thicksim 20$ $\mu$m). Figure \ref{Fig6} describes the observed scenarios for pair interactions. We distinguish three possible configurations depending on particle wetting configurations and the texture of the smectic around them: First, when two particles with coron$\ae$ approach each other (Fig.\ref{Fig6}-a), second when one particle with the corona approaches another particle without the corona (Fig.\ref{Fig6}-b), and finally when two particles without coron$\ae$ approach each other (Fig.\ref{Fig6}-c). In all scenarios, we see strong local capillary interactions in the near field, when the menisci on the colloids overlap (see figures SM Fig.2 and SM Fig.3 in the supplementary materials). We note that this is an unusual capillary interaction, as menisci around particles trapped in thin films of isotropic fluids decay over very long ranges, whereas these stop abruptly, limiting the range of capillary interaction to a well-defined separation distance. Moreover, particles also interact via elastic interactions, as we see relaxation of the smectic layers as the menisci merge or as the particles snap into the pore meniscus.

In the first case, of two particles with coron\ae, the particles interact in a peculiar way at a distance of around  $d_1  \thicksim 15$ $\mu$m; the particles' attraction is balanced by a repulsive force at this separation, so the particles diffuse weakly about $d=d_1$ for several minutes, before they suddenly come into contact. 
We attribute these complex dynamics to the pair of coron$\ae$ and the trapped oily streaks within them, which give an effectively steric repulsion (Fig.\ref{Fig6}-a). This behavior is similar to observations reported in previous studies on colloidal dispersions in nematic and smectic LCs \cite{ref18,ref32,ref51}.  It indicates that the interparticle separation at $d_1$ is not stable, because of the strong capillary attraction at short distances. The jump from $d_1  \thicksim 15 \mu$m to the particles coming into contact requires the FCDs between the two particles to disappear or rearrange (Fig.\ref{Fig6}-d); this significant reorganization implies an important energy barrier. 

In the second case, when one particle has a corona but the other one does not, we find related behavior. Particles fix their distance at $d_2  \thicksim 12$ $\mu$m, diffuse for a few minutes, then come into contact. However, the typical metastable distance $d_2  \thicksim 12$ $\mu$m in this case is smaller than the previous configuration $d_1  \thicksim 15$ $\mu$m  (Fig.\ref{Fig6}-b). We attribute this shorter-ranged repulsion to the fact that only one particle is surrounded by oily streaks. Finally, when two particles without coron$\ae$ approach each other, there is no steric-like repulsion at a metastable separation distance because there are no FCDs.  The particles simply attract each other via smectic elasticity and the strong capillary interaction in the near field until they come into contact (Fig.\ref{Fig6}-c).

This interplay between elasticity and capillarity is therefore at the origin of the novel structures that we observe in our system shown in Fig.\ref{Fig6}-d. These interactions seem to be anisotropic for multi-particle clusters. For example, after the formation of dimers, the latter rotate to orient in a specific manner. This behavior is clearly observed in and explains the formation of chains shown in Fig.\ref{Fig6}-d. These chains tend to orient themselves along a direction perpendicular to the radial lines of the circular film, and continue to collect more particles. For large aggregates, we find that some of them tend to form immediately after the deposition of particles (beads close to each other attract via strong capillary forces). These aggregates move at the free-standing film and collect other particles to form the patterns shown in Fig.\ref{Fig6}-d. The resulting structures range from 1D chains to 2D aggregates that may or may not be decorated by FCDs depending on the initial texture of the smectic around the isolated beads and their wetting configurations. Where FCDs are present, they surround the colloidal aggregate regardless of its shape. These resulting structures are the product of the strong colloidal interactions at smectic interfaces that offer new degrees of freedom to direct colloidal assembly. Eventually the flow continues to bring the colloids and their aggregates close enough to snap into the meniscus at the edge of the pore. There, expelled from the thin film, they do not distort the interface strongly and minimize the capillary energy.

\section{Conclusion}

We report novel colloidal interactions and assemblies generated by embedding solid particles at free-standing smectic-A films. We have described the role of elastocapillarity in controlling the behavior of particles at smectic interfaces. Particles captured in thin smectic films with different wetting conditions induce distortions of the smectic interface and lead to capillary forces between particles. Also, particles create bulk deformations of the smectic layers that give rise to elastic forces. The resulting potential induced by the interplay between elasticity and capillarity drives the assembly of particles into novel structures ranging from 1D chains to 2D aggregates. Coron$\ae$ of focal conic domains enhance the metastability of these unusual structures when the particles are trapped at one interface. We  believe this work paves the way to explore new means to manipulate ordering transitions at LC interfaces and suggests new strategies for making novel optically active materials. This also provides a novel route to making hierarchical functional structures, as FCDs have been used to organize nanoparticles within their cores \cite{ref54, ref55}.

\begin{acknowledgments}
The authors acknowledge support from the National Science Foundation (NSF) through Materials Research Science and Engineering Centers Grant DMR-1120901 as well as the NSF Grant DMR-1262047.  M.A.G. acknowledges support from the Natural Sciences and Research Council (NSERC). This work was partially supported by a Simons Investigator Grant from the Simons Foundation (to R.D.K.).
\end{acknowledgments}


\section*{Supplementary Materials}

\begin{itemize}

\item \textbf{Fluorescence microscopy}

\item \textbf{The smectic meniscus profile}

\item \textbf{Pair interaction between particles at short distances}

%
%
%
%

\end{itemize}

\section*{Author Contributions}
All authors designed and performed the research, analyzed the data and wrote the article. The authors declare no competing financial interest. 


\pagebreak
\widetext
\begin{center}
\textbf{\large Supplementary Materials: \\ Elastocapillary driven assembly of particles at free-standing smectic-A films} \\  \smallskip
Mohamed Amine Gharbi, Daniel A.\ Beller, Nima Sharifi-Mood, Rohini Gupta, \\ Randall D.\ Kamien, Shu Yang, and Kathleen J.\ Stebe
\end{center}
\setcounter{equation}{0}
\setcounter{figure}{0}
\setcounter{table}{0}
\setcounter{page}{1}
\setcounter{section}{0}
\makeatletter
\renewcommand{\theequation}{S\arabic{equation}}
\renewcommand{\thefigure}{S\arabic{figure}}
\renewcommand{\bibnumfmt}[1]{[S#1]}
\renewcommand{\citenumfont}[1]{S#1}

\section{Fluorescence microscopy}
We use the technique of fluorescence microscopy (FM) to verify the wetting configurations of particles at free-standing films. The fluorescent images shown in the SM Fig.1 are obtained using an inverted IX81 Olympus microscope equipped with an FV300 Olympus confocal scan box. The liquid crystal molecules are doped with a 0.01\% weight dye N,N′-Bis(2,5-di-tert-butylphenyl)-3,4,9,10-preylenedicarboximide (BTBP; Sigma–Aldrich). At these low concentrations, dye molecules align along the axis of the LC molecules while preserving the properties of 8CB and fluoresce when they are excited at 488nm. In the first case with the corona (SM Fig.1-a), we observe a dark circle at the upper interface and a bright region below the colloids, indicating the presence of smectic layers under them as shown in SM Fig.1-b and SM Fig.1-c. However, in the second case without the corona (SM Fig.1-d), the dark circles are found at both interfaces; the upper and lower ones (see SM Fig.1-e and SM Fig.1-f). The latter observations are in a good agreement with the polarized microscopy images and confirm the wetting configurations represented in Fig.4.

\begin{figure}[h]
\centering
\includegraphics[scale=0.7]{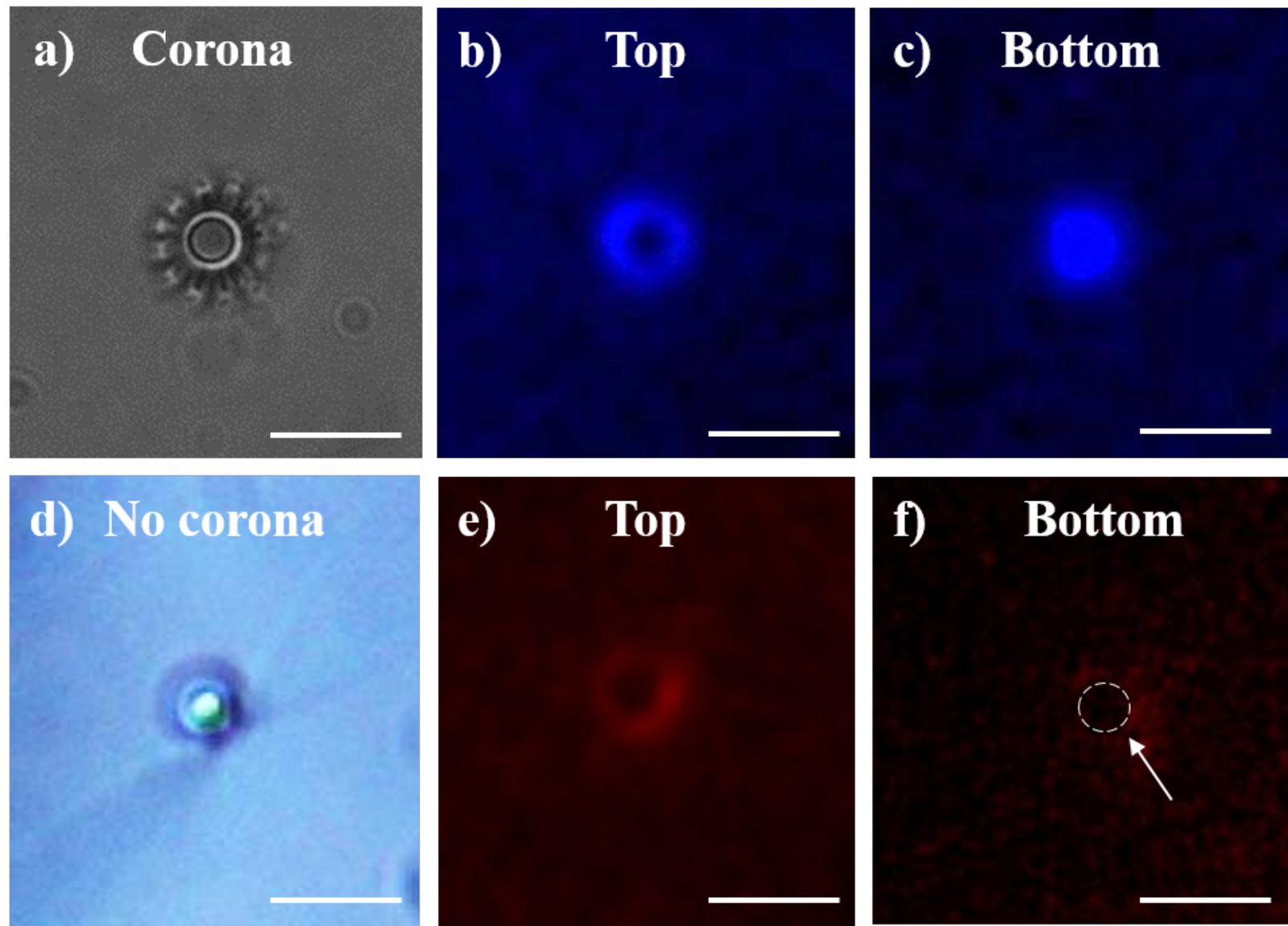}
\caption{Wetting configurations of particles at free-standing films verified with fluorescence microscopy. In the first case with the corona (a), we observe a dark circle at the upper interface (b) showing that the colloid is trapped at the interface. At the lower interface, we observe a bright spot below the colloid (c), indicating the presence of smectic layers. In the second case without the corona (d), the dark circles are observed at both, upper (e) and lower (f) interfaces, confirming that colloids are trapped by these interfaces. Scale bars: 15$\mu$m. }
\label{SMFig1}
\end{figure}

\section{The smectic meniscus profile}
Following Picano \textit{et al.} \cite{picano2000coupling}, we write the free energy for a smectic meniscus near a colloidal particle, approximated as a cylindrical wall, in the following form:

\begin{align}
F[h(r)] &= 2 \gamma_{SA} \int _{r_0}^{r_1} 2\pi r dr \left(\sqrt{1+(dh/dr)^2}-1\right) + 2 \left(\gamma_{PS}-\gamma_{PA}\right)2\pi r_0 h (r_0)  \nonumber \\
& \quad +2 \Delta p \int_{r_0}^{r_1} 2\pi r dr h(r) - \int_{r_0}^{r_1} 2\pi r dr E \frac{2}{a} \frac{dh}{dr} 
\label{Fpicano}
\end{align}
Here, $h(r)$ is the height profile as a function of radius $r$; the $\gamma$s are interfacial tensions where $S$ stands for smectic, $A$ for air, and $P$ for particle; $r_0$ is the inner radius of the meniscus; $r_1$ is defined by $h(r_1)=0$; $\Delta p$ is the pressure difference across the smectic membrane; $E$ is a defect core energy penalizing a density of dislocations; 
and $a$ is the layer thickness.

Because the flat membrane acts as a reservoir of particles, $\Delta p = 0$. The second term in Eqn. \ref{Fpicano} is determined by the boundary condition $h(r_0)$ set by the wetting chemistry. The Euler-Lagrange equation is then
\begin{align}
\frac{d}{dr} \left(\frac{r h'}{\sqrt{1+(h')^2}}\right) = e \label{PicanoEL}
\end{align}
where $e\equiv E/(\gamma_{SA}a)$ is a small parameter, and primes denote derivatives with respect to $r$. The solution at $e=0$ is the catenoid,
\begin{align}
h_c(r) &= h_\infty - r_\infty \ln \left[r/r_\infty +\sqrt{(r/r_\infty)^2-1}\right].  \label{catenoideqn}
\end{align}
For small but nonzero $e$, let $h(r) = h_c(r) + e \eta(r)$ where $e \eta(r) $ is a small correction to the meniscus profile. Expanding Equation \ref{PicanoEL} to first order in $e$, we have
\begin{align*}
e &\approx \frac{d}{dr} \left( \frac{r(h_c'+e \eta')}{\sqrt{1+(h_c')^2+2 e h'_c \eta' }}\right)   \\
&\approx \left\{\frac{d}{dr}\left(\frac{r h'_c}{\sqrt{1+(h'_c)^2}}\right)\right\} + e\frac{d}{dr}  \left [\frac{r\eta'}{\sqrt{1+(h'_c)^2}} - \frac{r (h'_c)^2\eta'}{\left(1+(h'_c)^2\right)^{3/2}}  \right]   \\
&= e \frac{d}{dr}  \left[\frac{r \eta'}{\left(1+(h'_c)^2\right)^{3/2}} \right ]
\end{align*}
where the term in curly braces vanishes because $h_c(r)$ satisfies Eqn.~\ref{PicanoEL} with $e=0$. Integration gives
\begin{align*}
r+c &=  \frac{r \eta'}{\left(1+(h_c')^2\right)^{3/2}}  \\
\Rightarrow \eta' &= \left(1+c/r\right) \left(1+(h'_c)^2\right)^{3/2} = \left(1+c/r\right) \left(1+\frac{1}{(r/r_\infty)^2-1}\right)^{3/2} \\
&= \left(1+c/r\right) \left(\frac{(r/r_\infty)^2}{(r/r_\infty)^2-1}\right)^{3/2}
\end{align*}
where $c$ is an integration constant. Integrating a second time gives
\begin{align}
\eta &= \sqrt{r^2-r_\infty^2} - \frac{r_\infty^2+c r}{\sqrt{r^2-r_\infty^2}}+ c \ln\left[(r/r_\infty)+\sqrt{(r/r_\infty)^2-1}\right] +c_2. \label{etasol} 
\end{align}
The third term is the catenoid again. The integration constant $c_2$ just gives an overall shift that can be incorporated into $h_\infty$, provided that $e$ has a fixed value. Adding this result to the catenoid profile of Eqn.~\ref{catenoideqn}, we obtain
\begin{align}
h(r) &= h_c(r) + e \eta (r) \nonumber \\
&= h_\infty +\left(ec - r_\infty \right) \ln \left[r/r_\infty +\sqrt{(r/r_\infty)^2-1}\right]  + e \left[ \sqrt{r^2-r_\infty^2} - \frac{r_\infty^2+c r}{\sqrt{r^2-r_\infty^2}} \label{hsol} \right].
\end{align}
Equation \ref{hsol} has four fitting parameters: $h_\infty$, $r_\infty$, $e$, and $c$. 

\section{Pair interaction between particles at short distances}

At the free-standing film, the circulating flow convects colloids towards neighboring particles before they are ejected to the meniscus. The colloids interact when
they come close enough to each other (center-to-center typical separation distance around $d \approx 20$ $\mu$m). However, in the near field, we see strong local capillary interactions. SM Fig.2 and SM Fig.3 show the latter interactions in the case of two particles with FCD coron$\ae$ and in the case of one particle with a corona interacting with a particle without a corona. We note that in SM Fig.2 and SM Fig.3, we don't observe a metastable equilibrium distance that we should obtain between particles before they attract each other. This is because the particles are initially trapped too close to each other, and the circulating flow causes them to approach each other.

\begin{figure}[h]
\centering
\includegraphics[scale=0.8]{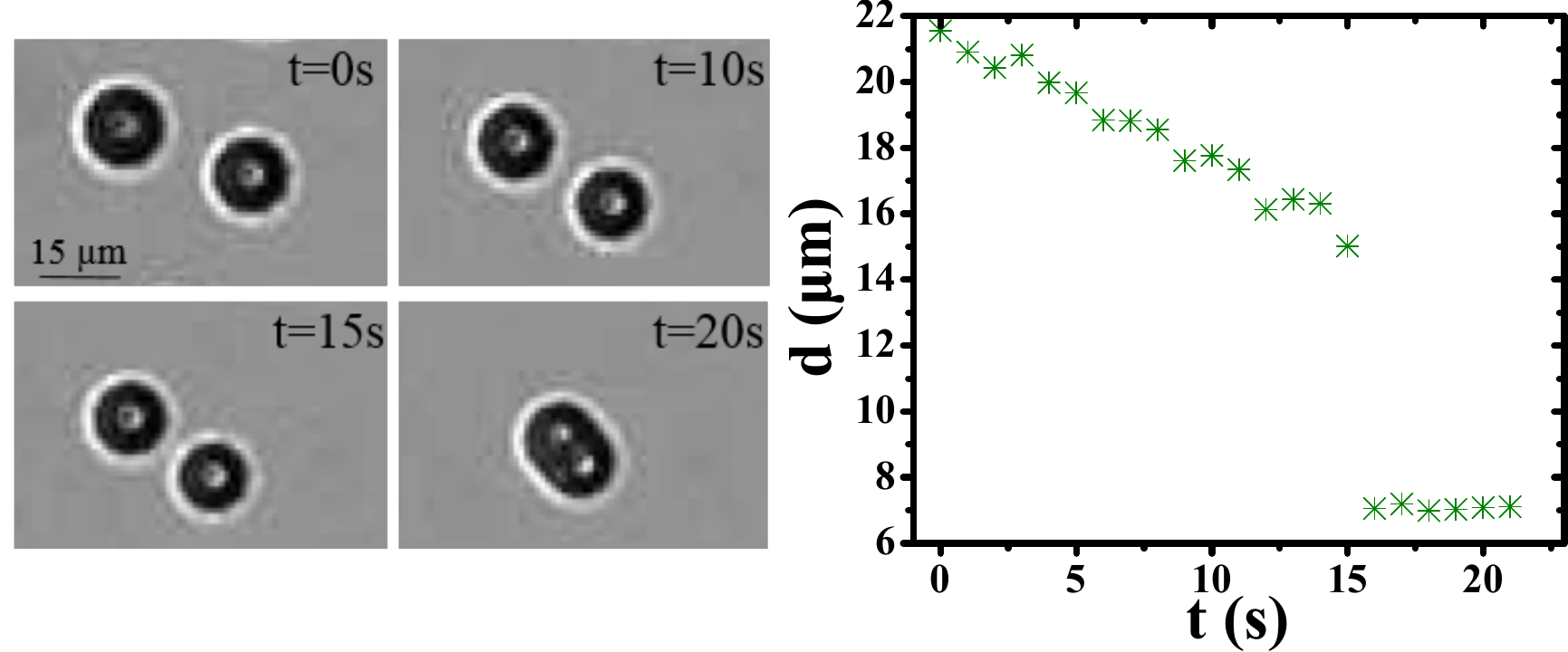}
\caption{Pair interaction at short distances for two particles with FCD coron$\ae$. Plot shows particle center-to-center separation as a function of time.}
\label{SMFig2}
\end{figure}

\begin{figure}[h]
\centering
\includegraphics[scale=0.8]{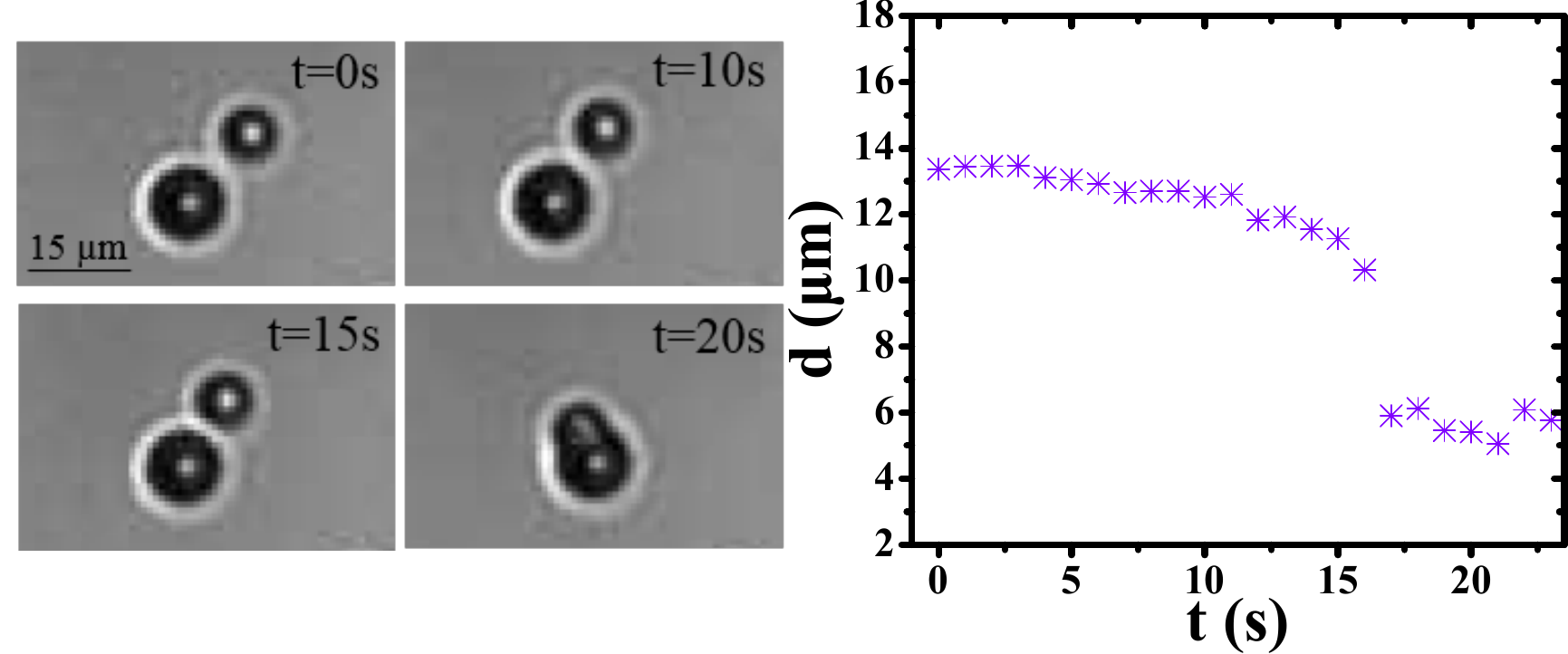}
\caption{Pair interaction at short distances between a particle with a corona and a particle without a corona. Plot shows particle center-to-center separation as a function of time.}
\label{SMFig2}
\end{figure}

%
%
%
%
%
%
%

\bibliographystyle{plain}

\end{document}